# The effect of oxygen stoichiometry on electrical transport and magnetic properties of La$_{0.9}$Te$_{0.1}$MnO$_y$

J. Yang [1], W. H. Song [1], R. L. Zhang [1], Y. Q. Ma [1], B. C. Zhao [1], Z. G. Sheng [1], G. H. Zheng [1], J. M. Dai [1] and Y. P. Sun [1,2,3]

[1] Key Laboratory of Materials Physics, Institute of Solid State Physics, Chinese Academy of Sciences, Hefei, 230031, P. R. China

[2] National Laboratory of Solid State Microstructures, Nanjing University, Nanjing 210008, P. R. China

**Abstract**

The effect of the variation of oxygen content on structural, magnetic and transport properties in the electron-doped manganites La$_{0.9}$Te$_{0.1}$MnO$_y$ has been investigated. All samples show a rhombohedral structure with the space group $R\bar{3}C$. The Curie temperature $T_C$ decreases and the paramagnetic-ferromagnetic (PM-FM) transition becomes broader with the reduction of oxygen content. The resistivity of the annealed samples increases slightly with a small reduction of oxygen content. Further reduction in the oxygen content, the resistivity maximum increases by six orders of magnitude compared with that of the as-prepared sample, and the ρ(T) curves of samples with y = 2.86 and y = 2.83 display the semiconducting behavior ($d\rho/dT < 0$) in both high-temperature PM phase and low-temperature FM phase, which is considered to be related to the appearance of superexchange ferromagnetism (SFM) and the localization of carriers. The results are discussed in terms of the combined effects of the increase in the Mn$^{2+}$/(Mn$^{2+}$+Mn$^{3+}$) ratio, the partial destruction of double exchange (DE) interaction, and the localization of carriers due to the introduction of oxygen vacancies in the Mn-O-Mn network.

The hole-doped magnanites $Ln_{1-x}A_xMnO_3$ (Ln = La-Tb, and A = Ca, Sr, Ba, Pb, etc.) have attracted much renewed attention because of their peculiar electrical transport and magnetic properties, especially the property of colossal magnetoresistance (CMR), which is promising for their potential applications such as magnetic reading heads, field sensors and memories [1-3]. All these researches indeed suggested that the ratio of $Mn^{3+}/Mn^{4+}$ is a key component for understanding the colossal magnetoresistance (CMR) effect and the transition from the ferromagnetic (FM) metal to paramagnetic (PM) semiconductor. The change in the ratio of $Mn^{3+}/Mn^{4+}$ can be achieved by varying the doping level of the divalent elements. However, the variation in the ratio of $Mn^{3+}/Mn^{4+}$ can be produced by the creation of vacancies at the oxygen sites as well. It was reported that the oxygen deficiency had dramatic effect on the electrical transport and magnetic properties of hole-doped manganites [4-12].

Recently, based on the consideration of the development of spintronics, many researches have been focused on electron-doped compounds such as $La_{1-x}Ce_xMnO_3$ [13-16], $La_{1-x}Zr_xMnO_3$ [17], $La_{2.3-x}Y_xCa_{0.7}Mn_2O_7$ [18], and $La_{1-x}Te_xMnO_3$ [19-21]. These investigations also suggest that the CMR behavior probably occur in the mixed-valent state of $Mn^{2+}/Mn^{3+}$. The basic physics in terms of Hund's rule coupling between $e_g$ electrons and $t_{2g}$ core electrons and Jahn-Teller (JT) effect due to $Mn^{3+}$ JT ions can operate in the electron-doped manganites as well. What does the role of vacancies at the oxygen sites play in electron-doped manganites? To our best knowledge, the relevant report is very few. To understand comprehensively the origin of the effect of oxygen vacancies on physical properties of manganites, in this article, we will report our investigation of the effect of oxygen vacancies on structural, magnetic and electrical transport properties in the electron-doped manganites $La_{0.9}Te_{0.1}MnO_y$.

Polycrystalline $La_{0.9}Te_{0.1}MnO_3$ samples were prepared through the conventional solid-state reaction method in air. Appropriate proportions of high-purity $La_2O_3$, $TeO_2$, and MnO$_2$ powders were thoroughly mixed according to the desired stoichiometry, and then calcined at 700 °C for 24h. The powders obtained were ground, pelletized, and sintered at 1030°C for 96h with three intermediate grindings, and finally, the furnace was cooled down to room temperature. In order to change the oxygen content of the samples, the samples were annealed at 750 °C, 800 °C and 850 °C, respectively, in N$_2$ atmosphere under 2MPa pressure for 4h with graphite powder placed near the samples. The oxygen content of the samples was determined by a redox (oxidation reduction) titration in which the powder samples taken in a quartz crucible were dissolved in (1+1) sulfuric acid containing an access of sodium oxalate, and the excess sodium oxalate was titrated with permanganate standard solution. The method is found to be effective and highly reproducible. The detailed method to determine the oxygen content of samples will be reported in elsewhere [22].

The crystal structures were examined by x-ray diffractometer using Cu $K_\alpha$ radiation at room temperature. The magnetic measurements were carried out with a Quantum Design superconducting quantum interference device (SQUID) MPMS system (2 $\leq T \leq$ 400 K, 0 $\leq$ H $\leq$ 5 T). Both zero-field-cooling (ZFC) and field-cooling (FC) data were recorded. The resistance was measured by the standard four-probe method from 25 to 300 K.

Fig.1 shows the x-ray diffraction (XRD) patterns of La$_{0.9}$Te$_{0.1}$MnO$_y$ samples of (A) as-prepared and (B) 750 °C, (C) 800 °C, (D) 850 °C-annealed in N$_2$ atmosphere. It exhibits that the x-ray diffraction pattern of all the annealed samples is similar to that of the as-prepared sample, implying that no structure change occurs, although the oxygen content of the annealed samples is reduced remarkably as shown below. The oxygen content of samples is determined to be 3.01, 2.97, 2.86 and 2.83 corresponding to samples A, B, C and D, respectively. It indicates that the oxygen stoichiometry decreases with increasing the annealed temperature, which is consistent with the results reported in hole-doped manganites [9-10]. In addition, the powder x-ray diffraction at room temperature shows that all samples are single phase with no detectable secondary phases and the samples can be indexed with the rhombohedral structure with the space group $R\bar{3}C$. The structural parameters of the samples are refined by the standard Rietveld technique [23]. Figs. 2(a) and 2(b) show the experimental and calculated XRD patterns for samples A and B, respectively. It shows that the fitting between the experimental spectra and the calculated values is quite well. The obtained structural parameters are listed in Table 1. It shows that the lattice parameters of $La_{0.9}Te_{0.1}MnO_y$ samples vary monotonically with decreasing oxygen content. The Mn-O-Mn bond angle decreases and the Mn-O bond length increases with the reduction of oxygen content. It is mainly related to the local lattice distortion caused by introducing oxygen vacancies in the Mn-O-Mn network. Moreover, the unit cell volume of samples increases monotonically with reducing the oxygen content. With the reduction of oxygen content, the average manganese oxidation state decreases and thus the average manganese ionic size increases resulting in the increase of unit cell volume of the samples.

Fig.3 shows the temperature dependence of magnetization M of $La_9Te_{0.1}MnO_y$ under both zero-field cooling (ZFC) and field cooling (FC) modes at H = 0.1 T for all samples. The Curie temperature $T_C$ (defined as the one corresponding to the peak of $dM/dT$ in the M vs. T curve) is 239 K, 238 K, 213K and 165K for samples A, B, C and D, respectively. Obviously, the Curie temperature $T_C$ of sample B decreases slightly compared with that of sample A. However, the magnetization magnitude of sample B decreases obviously at low temperatures compared with that of sample A. With further reducing the oxygen content, $T_C$ of sample C and D drop rapidly compared with that of the as-prepared sample. We suggest that the $T_C$ reduction should be attributed to the effect of the partial destruction of $Mn^{2+}$-O-$Mn^{3+}$ DE interaction network because of the introduction of the oxygen vacancies and the weakening of $Mn^{2+}$-O-$Mn^{3+}$ DE interaction arising from the decrease of the bandwidth of $e_g$ electrons due to the increase of Mn-O bond length and the decrease of Mn-O-Mn bond angle. In addition, Fig.3 shows that both samples A and B have a sharp PM to FM transition. However, compared with samples A and B, PM-FM phase transitions of samples C and D obviously become broader, which implies the increase of magnetic inhomogeneity with reducing oxygen content. It is worth noting that a bifurcation begins to occur between both FC and ZFC curves at low temperature region for samples B, C and D implying the appearance of magnetic frustration due to the competition between FM and AFM exchange interaction. With the reduction of oxygen content, the difference between FC and ZFC curves becomes greater. For sample D, an obvious spin glass-like behavior appears, which is evidenced by a cusp in the ZFC curve at 100 K and a distinctive separation of the FC and ZFC curves. So the reduction of oxygen content could result in the decreasing of $T_C$ and the broadening of PM-FM transition.

The magnetization as a function of the applied magnetic field at 5K is shown in Fig.4. It shows that, for samples A, B and C, the magnetization reaches saturation at about 1T and keeps constant up to 5T, which is considered as a result of the rotation of the magnetic domain under the action of applied magnetic field. For sample D, the rapid increase of magnetization M (H) at low magnetic fields resembles ferromagnet with a long-range FM ordering corresponding to the rotation of magnetic domains, whereas the magnetization M increases continuously without saturation at higher fields, which reveals the existence of AFM phase corresponding to the linear variation of high-field region. It is well known that the competition between FM phase and AFM phase would lead to the appearance of the spin glass state. That is why the spin glass-like behavior exists in sample D. In addition, it is worth noting that the magnetization magnitude of sample C also increases slightly at low temperatures as H > 1 T compared with that of sample B. It is may be related to the excess reduction of oxygen content for sample C (y = 2.86) compared with that of sample B (y = 2.97). This phenomenon would be further explained below.

Fig.5 shows the temperature dependence of the inverse magnetic susceptibility $\chi_m$ for samples A, B, C and D. For ferromagnet, it is well known that in the PM region, the relation between $\chi_m$ and the temperature T should follow the Curie-Weiss law, i.e., $\chi_m = C/(T-\Theta)$, where $C$ is the Curie constant, and $\Theta$ is the Weiss temperature. The lines in Fig.5 are the calculated curves deduced from the Curie-Weiss equation. It can be seen from Fig.5 that the experimental curve in the whole PM temperature range is well described by the Curie-Weiss law. The Weiss temperature $\Theta$ is obtained to be 241K, 239K, 221K and 173K for samples A, B, C and D, respectively. For sample A and B, $\Theta$ values approach corresponding $T_C$ values. However, for sample C and D, $\Theta$ values are higher than corresponding $T_C$ values which may be related to the magnetic inhomgeneity. The Curie constant C deduced from the fitting data is 6.05, 5.76, 5.61 and 3.724K·cm$^3$/mol for samples A, B, C and D, respectively. And thus the effective magnetic moment μ$_{eff}$ can be obtained as 4.923, 4.799, 4.737 and 3.860 μ$_B$ for samples A, B, C and D, respectively. According to a mean field approximation [24], the theoretical effective magnetic moment μ$_{eff}$ can also be calculated as 4.98, 5.06, 5.29, and 5.35 μ$_B$ for samples A, B, C and D, respectively. The results indicates that the experimental value of the sample with oxygen deficiency is obviously lower than that expected theoretically. The difference may imply that increased Mn$^{2+}$ ions due to the reduction of oxygen content do not lie in high spin state.

Fig.6 shows the temperature dependence of resistivity ρ(T) for samples A, B, C and D at zero fields in the temperature range of 30-300K. For sample A, it shows that there exists an insulator-metal (I-M) transition at $T_{P1}$(= 246K) slightly higher than its $T_C$ (=239 K). In addition, there exists a shoulder at $T_{P2}$(=223 K) below $T_{P1}$, which is similar to the double peak behavior observed usually in alkaline-earth-metal-doped and alkali-metal-doped samples of LaMnO$_3$ [25-29]. Double peaks ($T_{P1}$=240K and $T_{P2}$=205 K) shift to lower temperatures for sample B. Compared with sample A, I-M transition at $T_{P1}$ becomes weak and the low-temperature resistivity peak at $T_{P2}$ becomes more obvious. It shows that the oxygen deficiency can substantially enhance the low-temperature resistivity peak at $T_{P2}$. This phenomenon is also observed in the literature [13,30]. Different from the origin of the $T_{P1}$ peak, the resistivity peak at $T_{P2}$ is believed to reflect the spin-dependent interfacial tunneling due to the difference in magnetic order between surface and core [27]. And this variation of double ρ peak behavior is presumably related to an increase both of the height and width of tunnel barriers with increasing oxygen deficiency. Moreover, the resistivity of sample B at low temperatures does not drop with decreasing temperature. Instead there is a slight upturn in ρ(T) at low temperatures, implying the appearance of carrier localization caused by the oxygen vacancies because of the reduction of oxygen content. The experimental data measured at applied field of 0.5 T for samples A and B in the temperature range of 30-300K are also recorded. For samples A and B, it can be seen from Fig.6 that the resistivity of samples decreases under the applied magnetic field, $T_{P1}$ peak position shifts to a higher temperature and $T_{P2}$ peak position does not nearly change, which also means the origin of $T_{P2}$ peak is different from that of $T_{P1}$ peak. As it can also be seen from Fig.7, for samples A and B, there are corresponding peaks in the vicinity of $T_{P1}$, which agrees with the DE model, whereas the corresponding peak at $T_{P2}$ is not observed. Here the MR is defined as $\Delta\rho/\rho_0 = (\rho_0 - \rho_H)/\rho_0$, where $\rho_0$ is the resistivity at zero field and $\rho_H$ is the resistivity at H = 0.5T. In addition, the reduction of the oxygen content decreases the MR effect and this case is obvious above 245K corresponding to the I-M transition at $T_{P1}$ of sample A. Moreover, the MR values increase with decreasing temperature, similar to MR behavior observed usually in polycrystalline samples of hole-doped manganites, which is considered to be related to grain boundaries [27,31]. So we consider the reason that the corresponding peak at $T_{P2}$ for samples A and B does not appear arises mainly from the MR value near the temperature $T_{P2}$ being small compared with the large low-temperature MR. And thus the corresponding peak at $T_{P2}$ is suppressed completely by the gradually ascending low-temperature MR.

For samples C and D, ρ(T) curves display the semiconducting behavior ($d\rho/dT < 0$) in both high-temperature PM phase and low-temperature FM phase and the resistivity maximum increases by six orders of magnitude compared with that of the as-prepared sample implying the enhancement of the localization of carriers. This FM insulating (FMI) phase was also found in $La_{1-x}Sr_xMnO_3$ [32-33] and $La_{1-x}Li_xMnO_3$ [28] compounds with orthorhombic structure. FMI behavior cannot be explained base only on the DE model because the FM and metallic nature must coexist within the framework of the DE model. Therefore the FM order at low temperatures for samples C and D does not mainly originate from DE interaction. As Goodenough predicts, a Mn-O-Mn 180°-superexchange interaction generally gives rise to AFM ordering while 90°-superexchange interaction will lead to FM ordering [34]. This superexchange ferromagnetism (SFM) has been suggested in explaining ferromagnetism of the $Tl_2Mn_2O_7$ and $CaCu_3Mn_4O_{12}$, which are of the character of no Mn mixed valence [35-36]. The Mn-O-Mn bond angel for sample C and D are 160.65° and 157.10°, which deviate obviously from 180°. Therefore, for samples C and D, the observed FM transition may be related to the appearance of SFM caused by the local lattice distortion arising from the increase of oxygen vacancies in the Mn-O-Mn network.

It should be mentioned that for $La_{0.9}Te_{0.1}MnO_{2.97}$ (sample B), the ratio of $Mn^{2+}/(Mn^{2+}+Mn^{3+})$ is close to 16%, comparable to that of $La_{0.84}Te_{0.16}MnO_3$. From the known experimental data [19], such a system should show higher I-M transition temperature and lower resistivity compared with that of sample A. For $La_{0.9}Te_{0.1}MnO_{2.86}$ (sample C) and $La_{0.9}Te_{0.1}MnO_{2.83}$ (sample D), the ratio of $Mn^{2+}/(Mn^{2+}+Mn^{3+})$ is close to 38% and 44%, comparable to that of $La_{0.62}Te_{0.38}MnO_3$ and $La_{0.56}Te_{0.44}MnO_3$, respectively. In our previous work [37], such two systems should show an I-M transition. However, the case is clearly not observed in samples C and D. Additionally, as we can see from Fig.4, the magnetization magnitude of sample C at low temperatures as H > 1 T is slightly higher than that of sample B, which is consistent with the experimental data due to the increase of the ratio of $Mn^{2+}/(Mn^{2+}+Mn^{3+})$[37]. On the other hand, for sample C, the appearance of SFM caused by the local lattice distortion arising from the increase of oxygen vacancies in the Mn-O-Mn network is also a possible reason. Moreover, the effect of oxygen content on $T_C$ of samples B, C and D is quite small comparable to that of $La_{0.84}Te_{0.16}MnO_3$, $La_{0.62}Te_{0.38}MnO_3$ and $La_{0.56}Te_{0.44}MnO_3$, respectively. So oxygen content reduction in $La_{0.9}Te_{0.1}MnO_y$ is expected to cause three effects. The first effect is the increase in the $Mn^{2+}/(Mn^{2+}+Mn^{3+})$ ratio, driving the carrier density increase and the resistivity decrease, and causing the enhancement of the FM coupling. The second effect is the partial destruction of double exchange (DE) interaction between $Mn^{2+}$-O-$Mn^{3+}$ and the localization of carriers because of the appearance of oxygen vacancies. The third effect is the occurrence of the local lattice distortion due to the introduction of oxygen vacancies in the Mn-O-Mn network, which is important for electrical conduction. The local lattice distortion caused by oxygen vacancies in samples is confirmed by the structural parameter fitting through the Reitveld technique. The local lattice distortion is also expected to increase the resistivity and destruct the DE interaction due to the reduced $e_g$ electron bandwidth [9-10]. As a result, the competitions of the three effects suggested above decide the complicated behavior of the electrical transport and magnetic properties of the samples $La_{0.9}Te_{0.1}MnO_y$.

In summary, we have studied the structural, electrical transport and magnetic properties of $La_{0.9}Te_{0.1}MnO_y$ with different oxygen contents. No structure change occurs with the increase of the oxygen deficiency and the Mn-O-Mn bond angle decreases with the reduction of oxygen content. The Curie temperature $T_C$ of samples decreases with increasing the oxygen deficiency. The resistivity of the samples increases slightly with small reduction of oxygen content. Further reduction of oxygen content leads to a dramatic increase in resistivity and ρ(T) curves of samples C and D display the semiconducting behavior ($d\rho/dT < 0$) in both high-temperature PM phase and low-temperature FM phase, which is considered to be originated from the appearance of superexchange ferromagnetism (SFM).

**ACKNOWLEDGMENTS**

This work was supported by the National Key Research under contract No.001CB610604, and the National Nature Science Foundation of China under contract No.10174085, Anhui Province NSF Grant No.03046201 and the Fundamental Bureau Chinese Academy of Sciences.

**TABLE 1. Refined structural parameters of $La_{0.9}Te_{0.1}MnO_y$ at room temperature. The space group is $R\bar{3}C$.**

| Sample | a (Å) | c (Å) | V (Å³) | $d_{Mn-O}$ (Å) | $\theta_{Mn-O-Mn}$ (°) | $R_p$ (%) |
|--------|-------|--------|---------|----------------|------------------------|-----------|
| A | 5.524 | 13.357 | 353.010 | 1.9644 | 163.83 | 8.03 |
| B | 5.533 | 13.367 | 354.407 | 1.9710 | 162.24 | 7.59 |
| C | 5.542 | 13.383 | 355.928 | 1.9783 | 160.65 | 8.72 |
| D | 5.549 | 13.396 | 356.859 | 1.9851 | 157.41 | 9.47 |

# Figure captions

Fig.1. X-ray diffraction patterns of the as-prepared $La_{0.9}Te_{0.1}MnO_y$ sample (denoted as A) and samples annealed at 750°C, 800°C and 850°C denoted as B, C and D, respectively.

Fig.2 The experimental and calculated XRD patterns of the compound $La_{0.9}Te_{0.1}MnO_y$, (a) sample A and (b) sample B. Crosses indicate the experimental data and the calculated data is the continuous line overlapping them. The lowest curve shows the difference between experimental and calculated patterns. The vertical bars indicate the expected reflection positions.

Fig.3. Magnetization as a function of temperature M (T) for samples A and B under the field-cooled (FC) and zero-field-cooled (ZFC) modes denoted as the filled and open symbols, respectively. The inset shows M (T) curves for samples C and D.

Fig.4. Field dependence of the magnetization M (H) for samples A, B, C and D at 5 K.

Fig.5. The temperature dependence of the inverse of the magnetic susceptibility for samples A, B, C and D. The lines are the calculated curves according to the Curie-Weiss law.

Fig.6. The temperature dependence of the resistivity $\rho(T)$ for samples A, B, C and D at zero (solid lines) and 0.5T fields (dashed lines).

Fig.7. The temperature dependence of magnetoresistance (MR) ratio for samples A and B at H = 0.5 T.

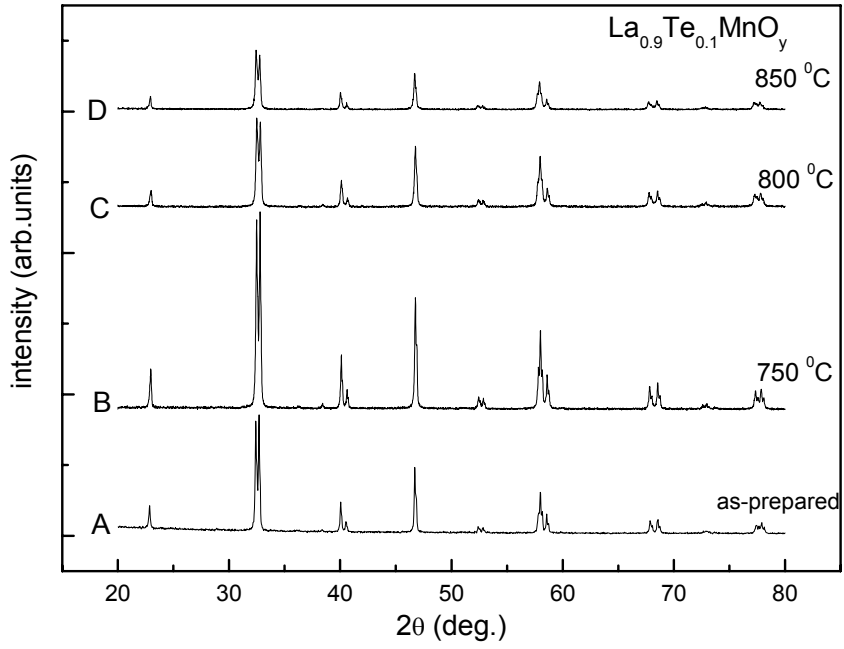

Figure.1 J.Yang et al.

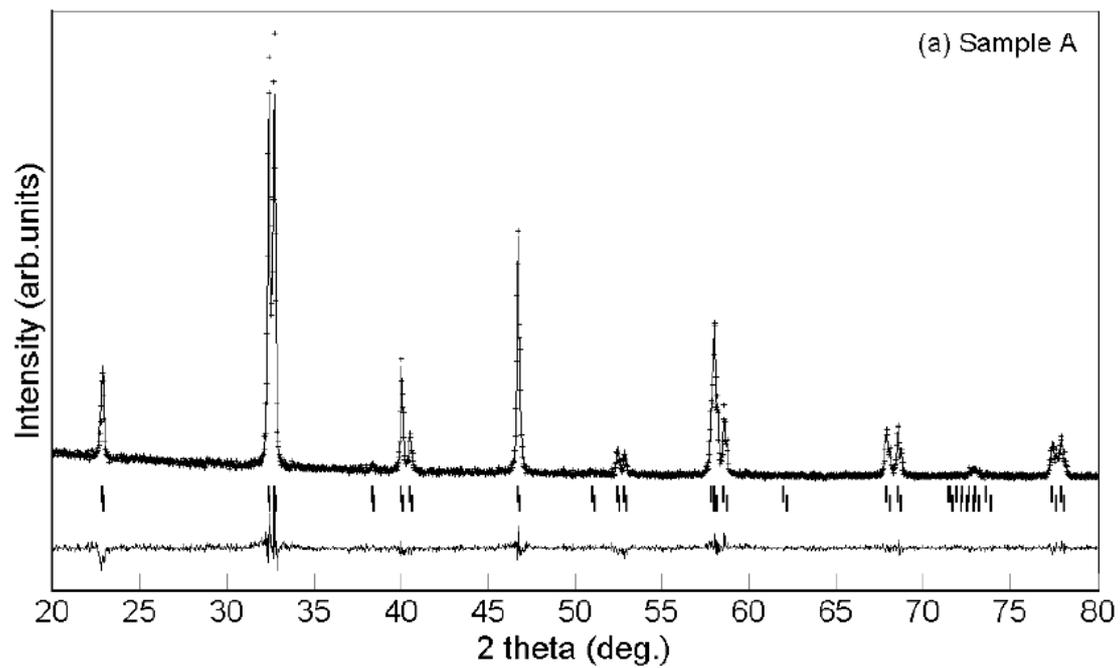

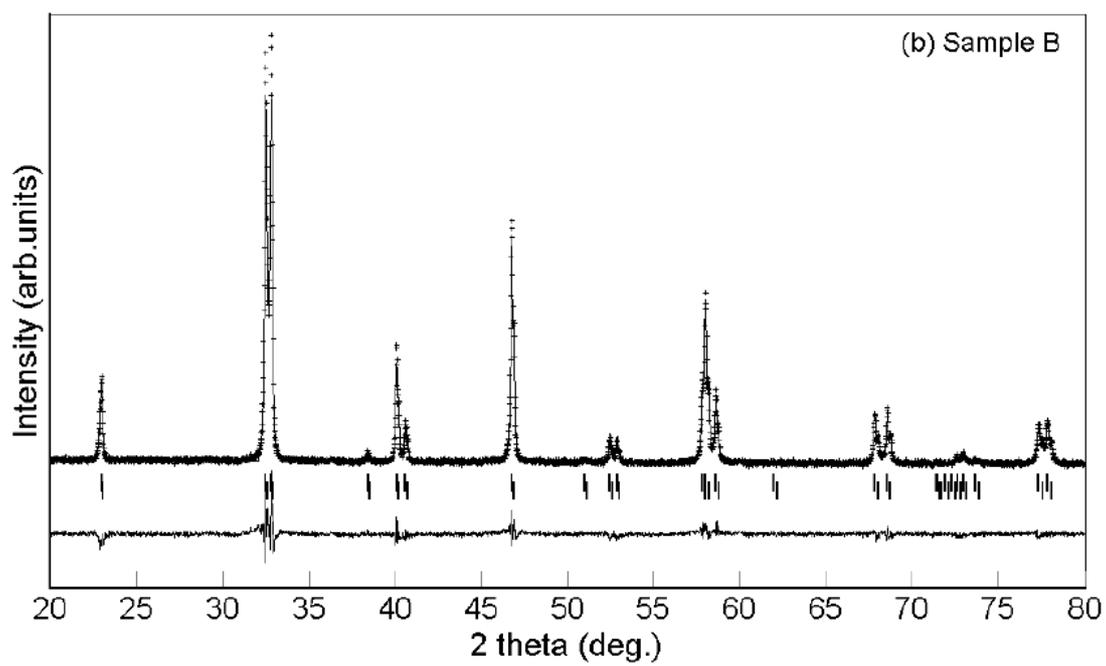

Fig.2 J. Yang et al.

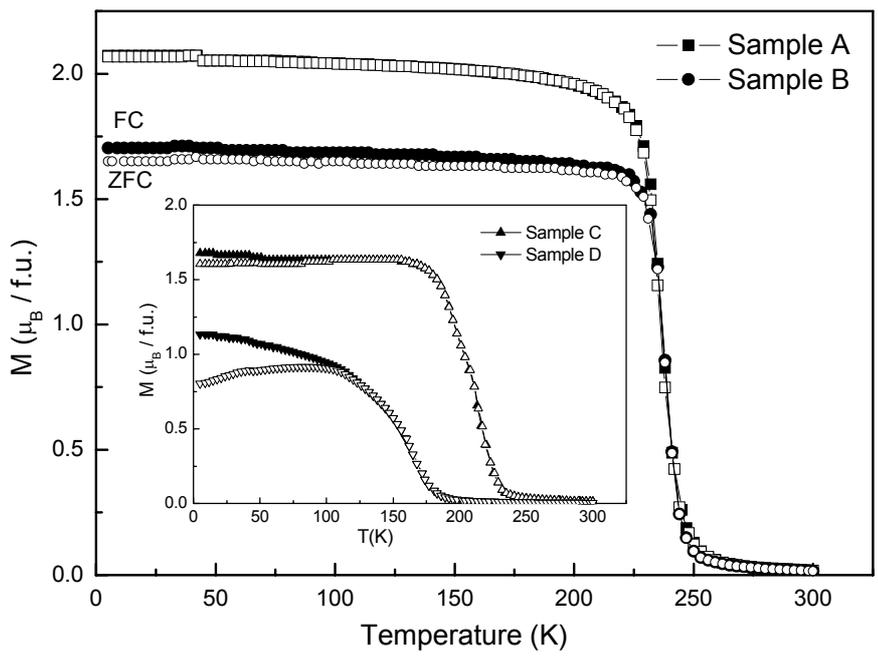

Figure.3 J.Yang et al.

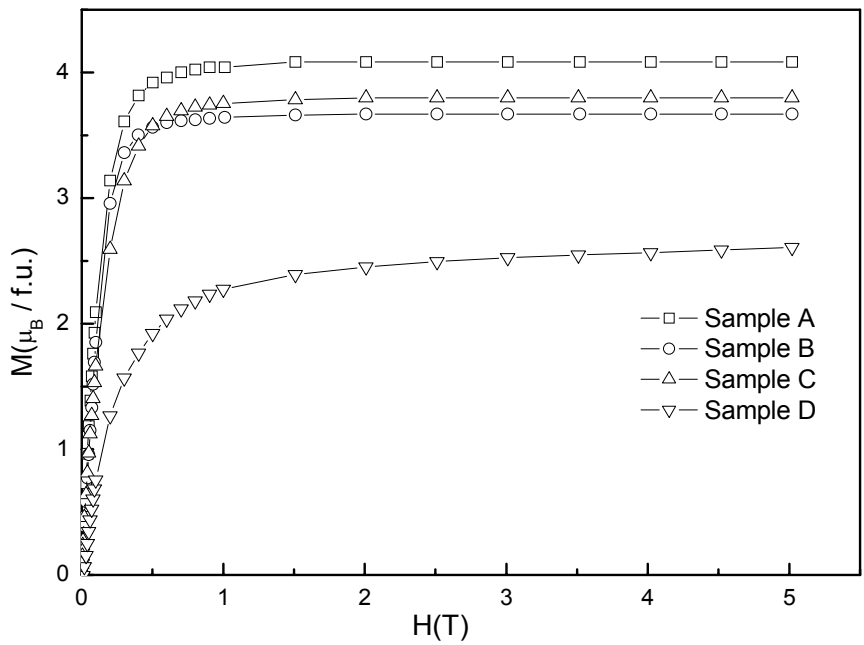

Figure.4 J.Yang et al.

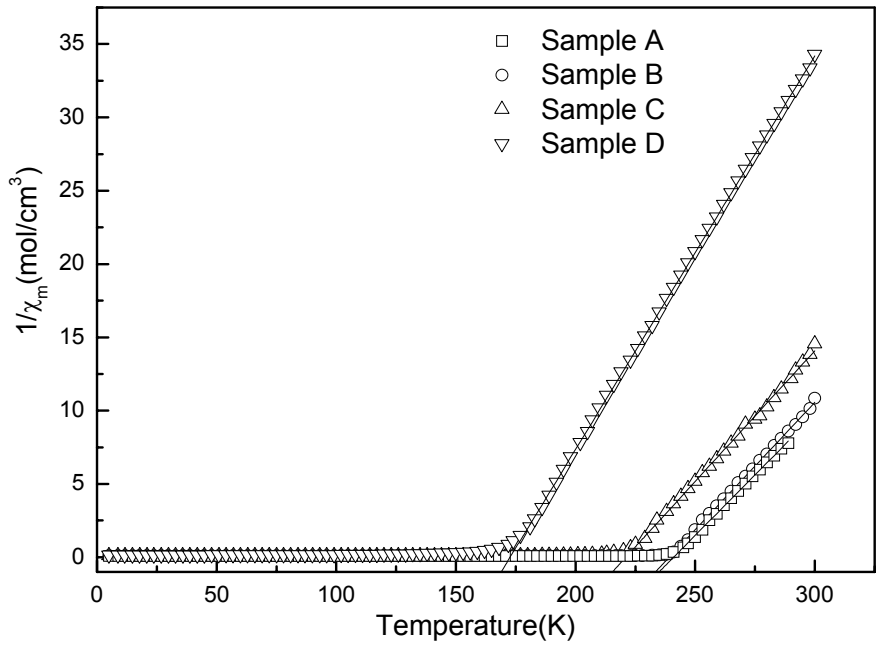

Figure.5 J.Yang et al.

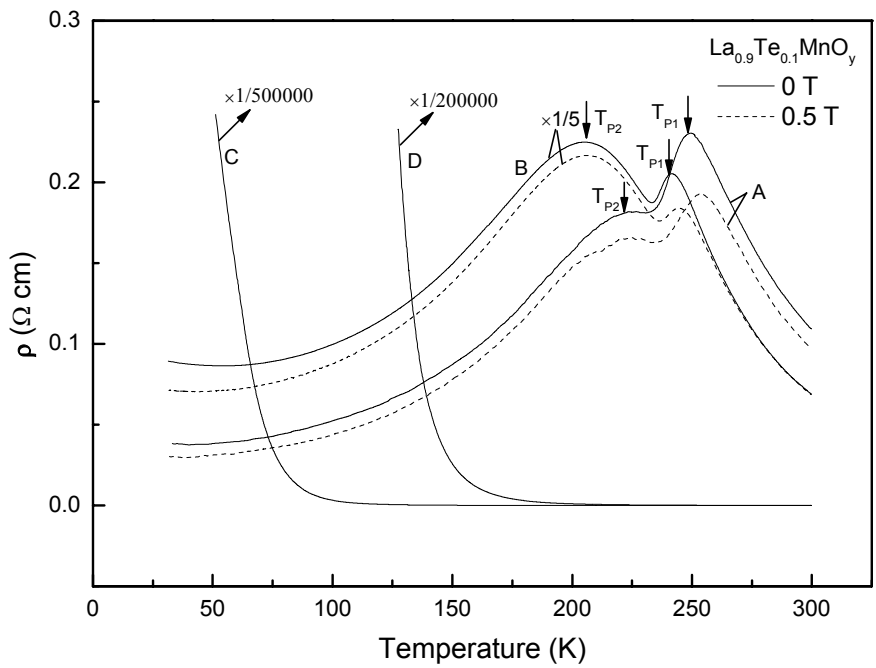

Figure.6 J.Yang et al.

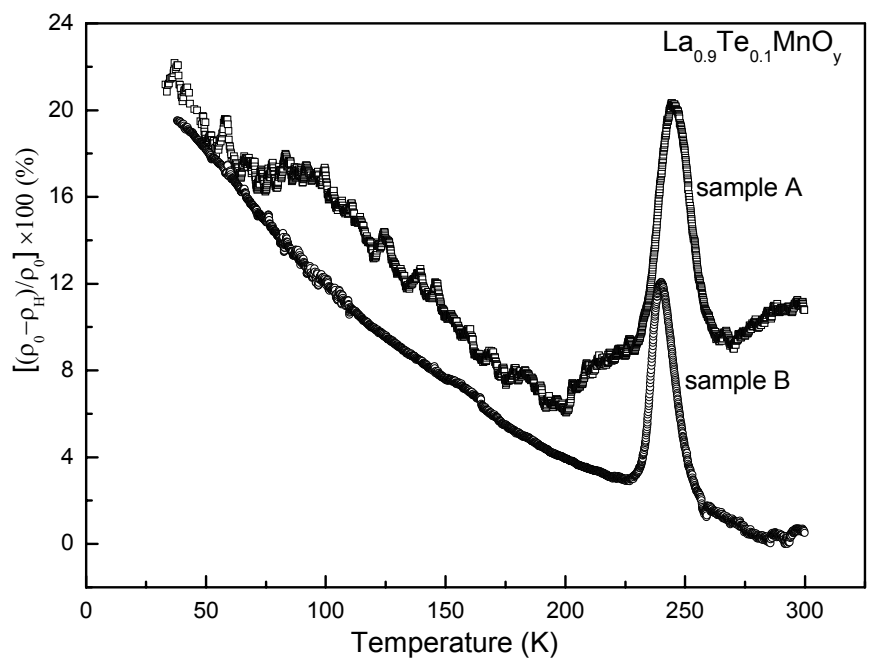

Figure.7 J.Yang et al.